\documentclass[pre,aps,twocolumn]{revtex4}
\usepackage{graphicx}

\begin{document}

\title{Three-Dimensional Nonlinear Lattices: From Oblique Vortices and
Octupoles to Discrete Diamonds and Vortex Cubes}
\author{R.\ Carretero-Gonz\'alez$^1$, P.G.\ Kevrekidis$^{2}$,
B.A.\ Malomed$^3$ and D.J.\ Frantzeskakis$^4$}
\affiliation{
$^1$ Nonlinear Dynamical Systems Group, 
Department of Mathematics \& Statistics,
San Diego State University, San Diego CA, 92182-7720 \\
$^{2}$ Department of Mathematics and Statistics,
University of
Massachusetts, Amherst MA 01003-4515, USA \\
$^3$ Department of Interdisciplinary Studies, Faculty of Engineering, Tel
Aviv University, Tel Aviv 69978, Israel \\
$^4$ Department of Physics, University of Athens, Panepistimiopolis,
Zografos, Athens 15784, Greece}

\begin{abstract}
We construct a variety of novel localized states with distinct
topological structures in the 3D discrete nonlinear
Schr{\"{o}}dinger equation. The states can be created in
Bose-Einstein condensates trapped in strong optical lattices, and
crystals built of microresonators. These new structures, most of
which have no counterparts in lower dimensions, range from purely
real patterns of dipole, quadrupole and octupole types to vortex
solutions, such as ``diagonal" and ``oblique" vortices, with axes
oriented along the respective directions $(1,1,1)$ and $(1,1,0)$.
Vortex ``cubes" (stacks of two quasi-planar vortices with like or
opposite polarities) and ``diamonds" (discrete skyrmions formed by
two vortices with orthogonal axes) are constructed too. We
identify stability regions of these 3D solutions and compare them
with their 2D counterparts, if any. An explanation for the
stability/instability of most solutions is proposed. The evolution
of unstable states is studied as well.
\end{abstract}

\maketitle

\textit{Introduction}. In the past two decades an explosion of
activity has been observed in the study of intrinsic localized
modes (ILMs, {\it alias} discrete solitons) in nonlinear dynamical
lattices, especially due to the ability of such modes to act as
energy ``hot spots" 
\cite{review}. The
relevance of ILMs has been demonstrated in problems ranging from 
arrays of nonlinear-optical waveguides \cite{mora} and photonic
crystals \cite{PhotCryst}, to Bose-Einstein condensates (BECs)
trapped in optical lattices (OLs) \cite{tromb} and
Josephson-junction ladders \cite{alex}.

A universal model, which may arise as an envelope approximation
from most of the complex nonlinear equations on the lattice and also
as direct physical model for the BECs \cite{tromb} and optical
waveguiding arrays \cite{fibers,fiberExper}, is the discrete
nonlinear Schr{\"{o}}dinger (DNLS) equation \cite{DNLS}. On top of
its significance to applications, the DNLS equation itself is a
fundamentally interesting dynamical model. In the 3D case, its
direct physical realization is provided, as mentioned above, by
BECs\ trapped in strong OLs \cite{tromb}. Waveguide arrays, however, cannot
be described by a 3D discrete model, since the evolution variable
in the optical media is a spatial coordinate, while the temporal
variable, which effectively plays the role of an additional
quasi-spatial one, cannot be discrete. Nevertheless, another physical
realization of the 3D DNLS equation may be provided by a crystal
built of microresonators \cite{photons}.

The study of the 3D continuum NLS equation, including a 3D
\cite{Salerno} or quasi-2D \cite{Salerno2} OL, and of the DNLS
model proper \cite{DNLS3d}, has started recently, becoming
increasingly accessible to numerical computations. As a result,
the first coherent structures, such as discrete vortices of the
topological charge (vorticity) $S=1,2$ and $3$, were identified
and their stability was investigated. The aim of the present work
is to study a large variety of novel localized 3D structures in
the DNLS equation, many of which turn out to be stable. In particular, 
we construct 
states
which include, first, dipoles with the axis oriented
along a lattice bond, or along a planar diagonal, or along a 3D
diagonal (we call them, respectively, ``straight", ``oblique", and
``diagonal" dipoles). Next, we construct quadrupole and octupole
states, that, similarly to the dipoles, are real solutions. More
sophisticated structures are also presented, namely ``vortex cubes" (concatenations of
two straight vortices with the same or opposite charges centered
on parallel planes), oblique and diagonal vortices, and ``vortex
diamonds", formed by a crossed pair of vortices with orthogonal
axes. 
Apart from the straight and oblique dipoles and quadrupoles,
these ILMs have no counterparts in 2D lattices.

To present the results, we first introduce the model, and then report
systematic numerical results for the shape and stability of the new
localized states. This is followed by conclusions, which include an
explanation for the stability and instability of the majority of patterns
found in this work.

\textit{The Model.} We consider the DNLS equation on the cubic lattice with
a coupling constant $C$ \cite{DNLS3d},
\begin{equation}
i\dot{\phi}_{l,m,n}+C\Delta \phi _{l,m,n}+\left\vert \phi
_{l,m,n}\right\vert ^{2}\phi _{l,m,n}=0,  
\label{NLS}
\end{equation}
with $\dot{\phi}=d\phi /dt$, and the discrete Laplacian is $\Delta \phi
_{l,m,n}\equiv \phi _{l+1,m,n}+\phi _{l,m+1,n}+\phi
_{l,m,n+1}+\phi _{l-1,m,n}+\phi _{l,m-1,n}+\phi _{l,m,n-1}-6\phi
_{l,m,n}$ . Solutions are looked for as $\phi
_{l,m,n}=u_{l,m,n}\exp (i\Lambda t)$ with a frequency $-\Lambda $
(or the chemical potential in the context of BEC), where the 
stationary functions $u_{l,m,n}$ obey the equation
\begin{equation}
\Lambda u_{l,m,n}=C\Delta u_{l,m,n}+\left\vert u_{l,m,n}\right\vert
^{2}u_{l,m,n}.  
\label{standing}
\end{equation}
Profiles used as an initial guess for the fixed-point iteration converging
to solutions displayed below, were based on the form of the
respective solutions (for the same $\Lambda $) in the
anti-continuum (AC) limit, $C=0$. Once solutions to Eq.\
(\ref{standing}) have been obtained, the linear-stability analysis
is performed for a perturbed solution \cite{DNLS3d}, $\phi
_{l,m,n}=\left[ u_{l,m,n}+\epsilon \left( a_{l,m,n}e^{\lambda
t}+b_{l,m,n}e^{\lambda ^{\ast }t}\right) \right] e^{i\Lambda t}$,
where $\epsilon $ is an infinitesimal amplitude of the
perturbation, and $\lambda $ is its eigenvalue. The Hamiltonian
nature of the system dictates that if $\lambda $ is an eigenvalue,
then so are $-\lambda $, $\lambda ^{\ast }$ and $-\lambda ^{\ast
}$(in the stable case, $\lambda $ is imaginary, hence this
symmetry yields only two different eigenvalues, $\lambda $ and
$-\lambda $). The stationary solution is unstable if at least one
pair of the eigenvalues features nonvanishing real parts.

\begin{figure}[tbp]
\begin{center}
\begin{tabular}{lll}
(a) & (b) & (c) \\[-2.5ex]
\hskip-0.2cm\includegraphics[width=2.95cm]{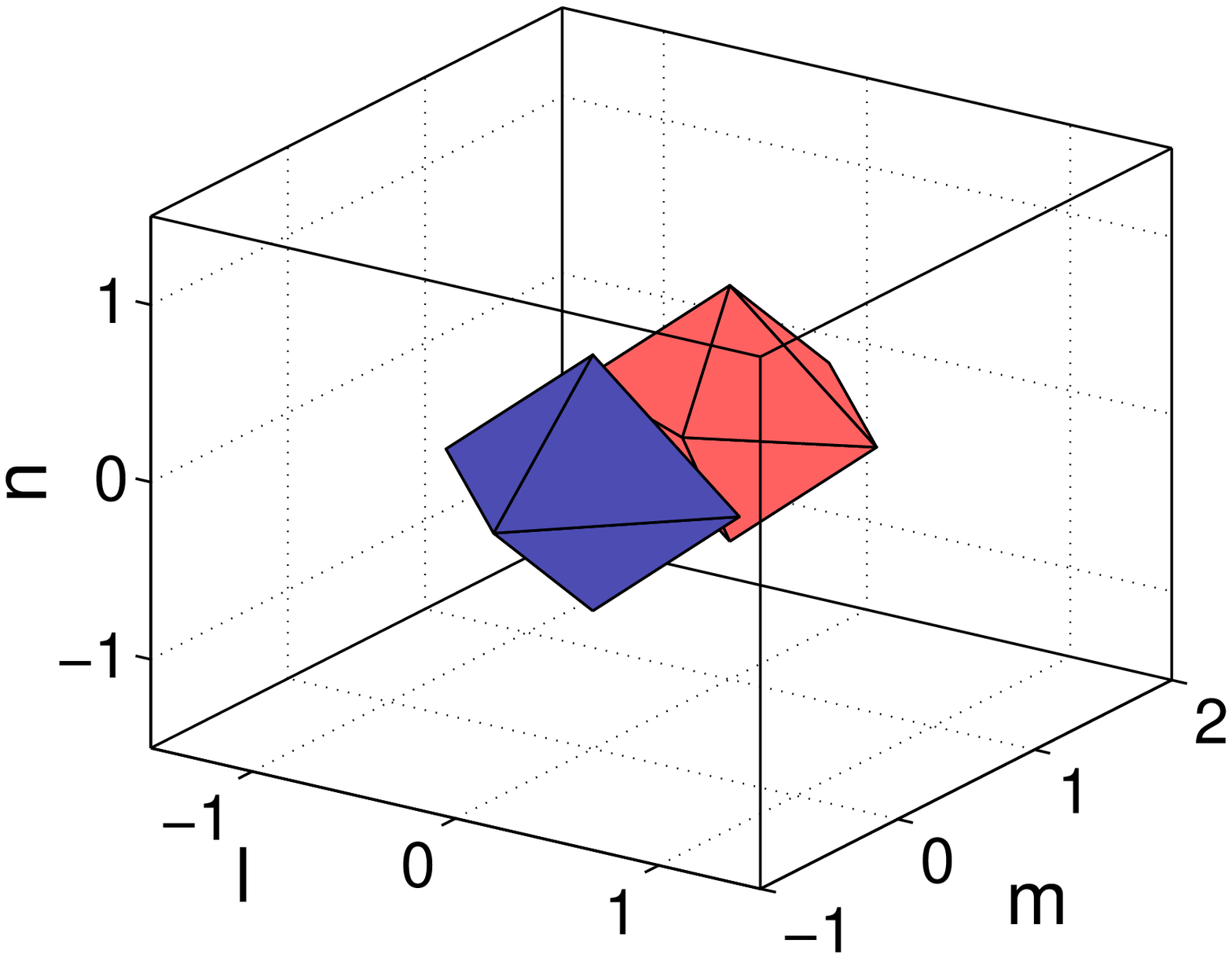} &
\hskip-0.2cm\includegraphics[width=2.95cm]{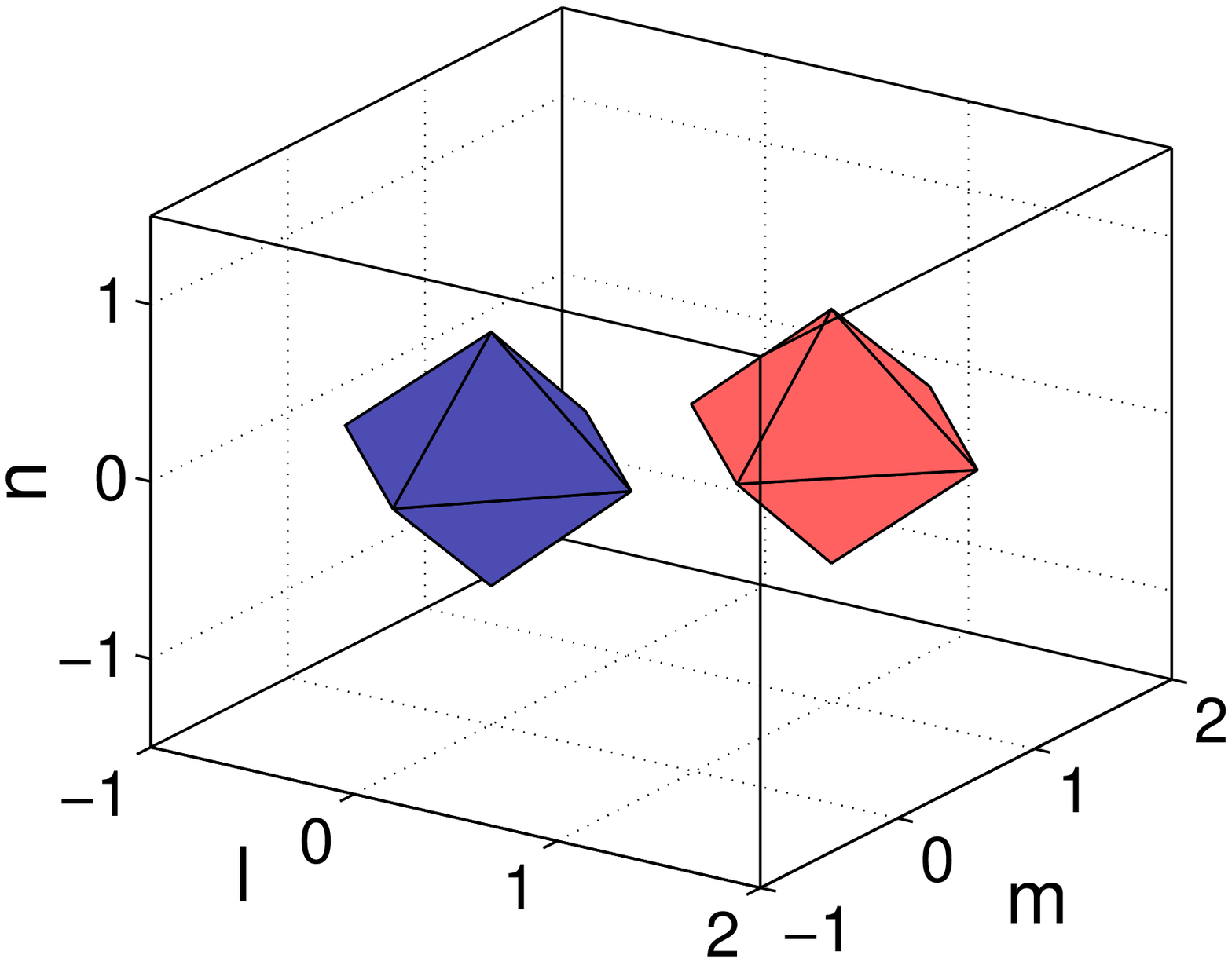} &
\hskip-0.2cm\includegraphics[width=2.95cm]{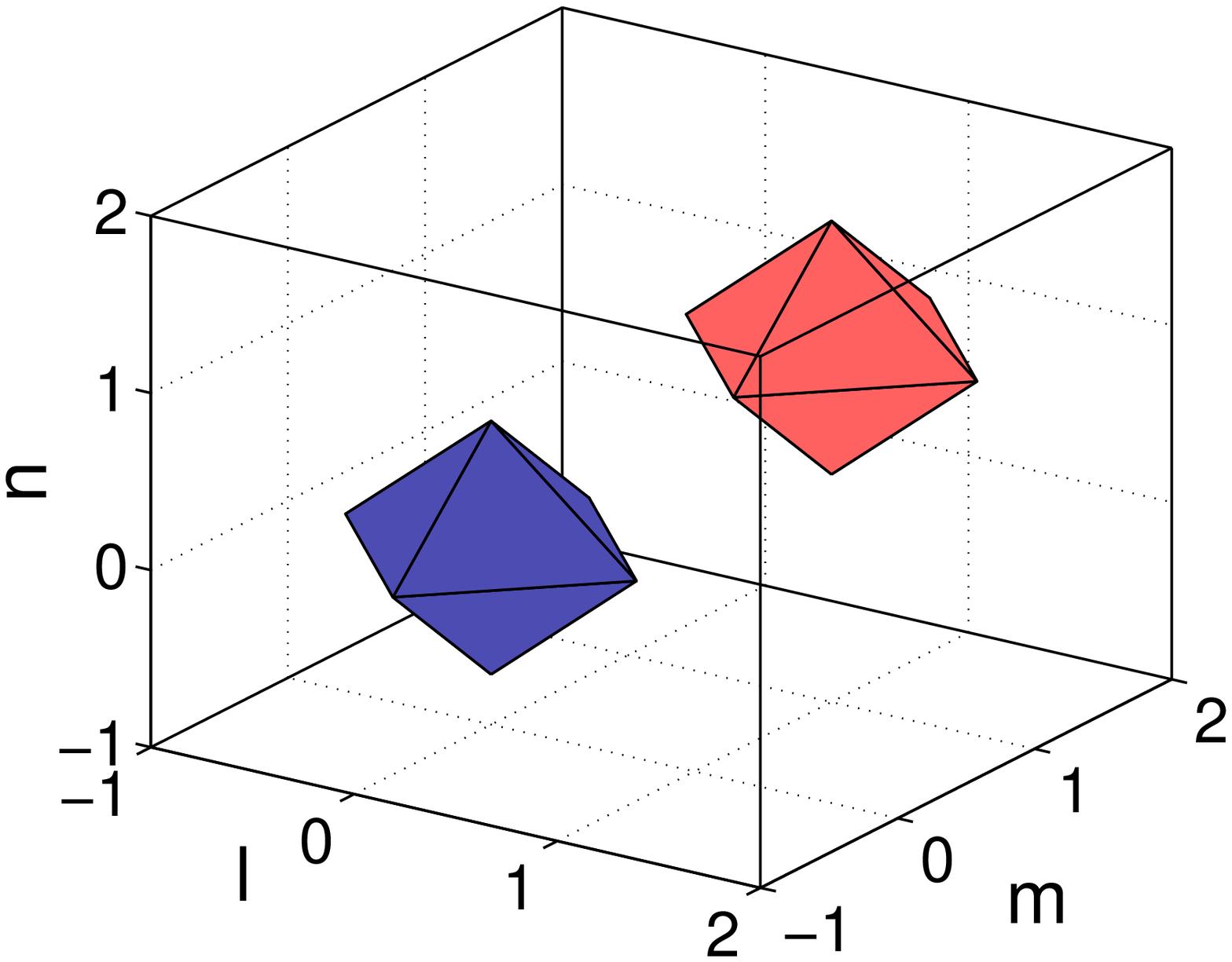} \\
(d) & (e) & (f) \\[-2.5ex]
\hskip-0.2cm\includegraphics[width=2.95cm]{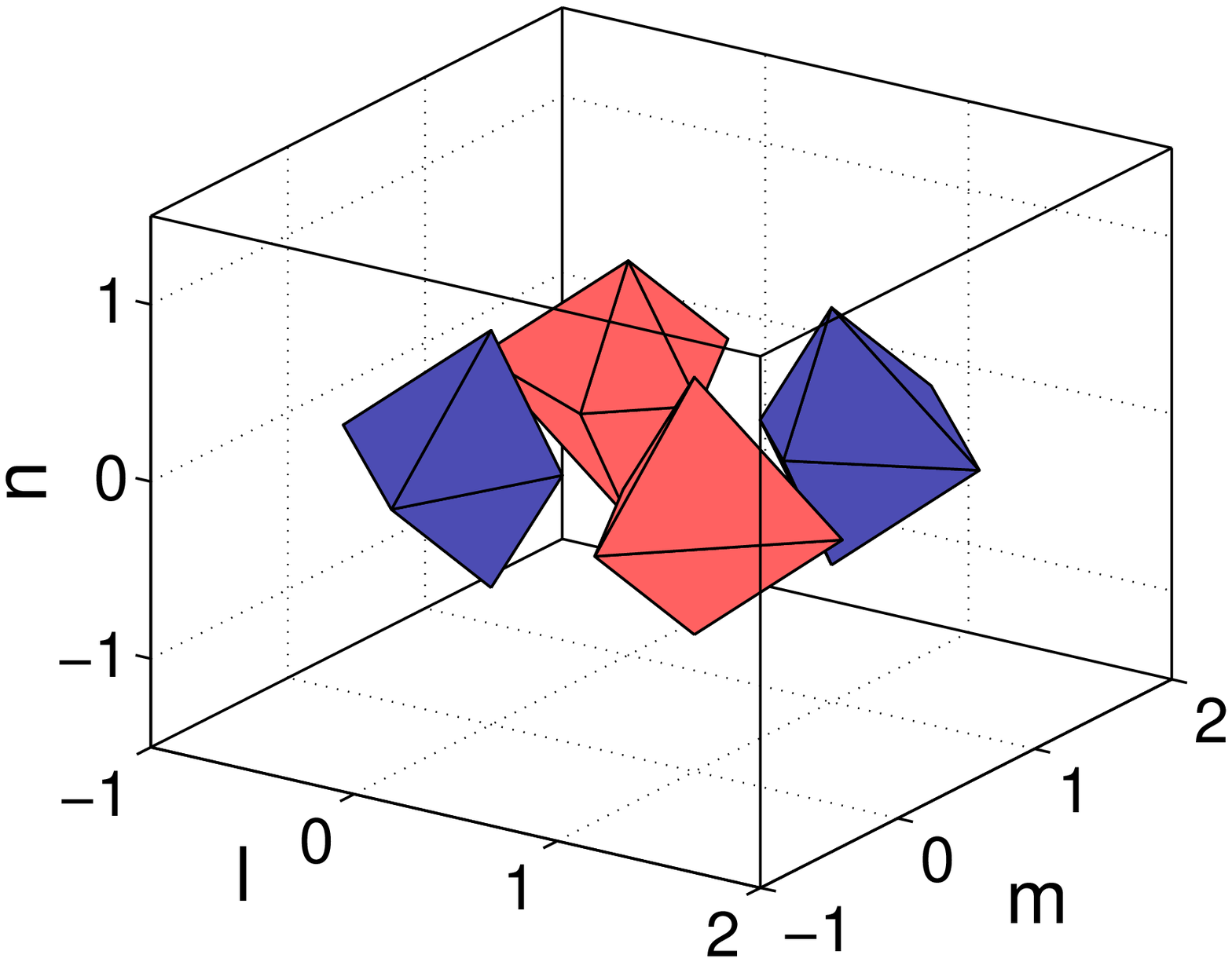} &
\hskip-0.2cm\includegraphics[width=2.95cm]{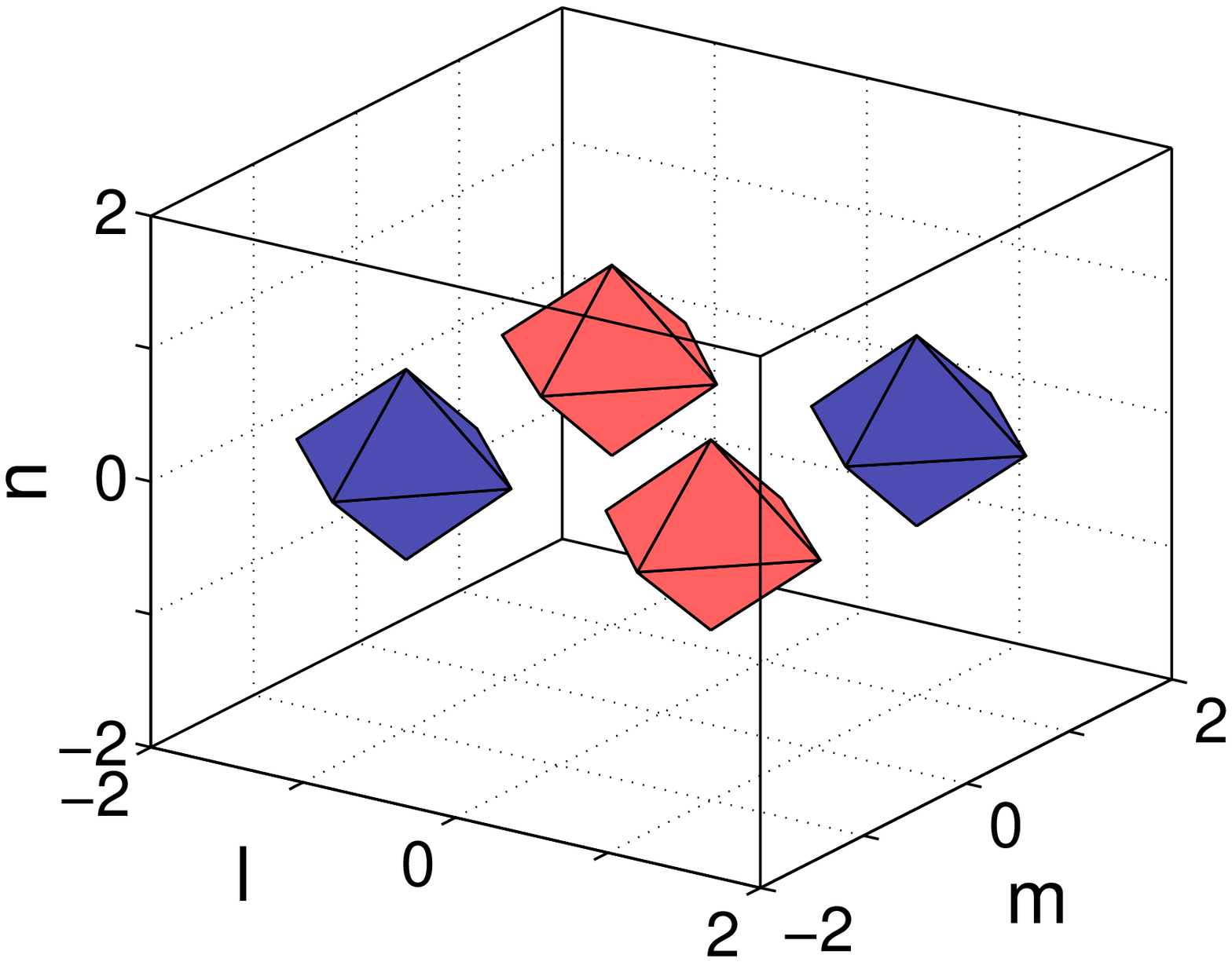} &
\hskip-0.2cm\includegraphics[width=2.95cm]{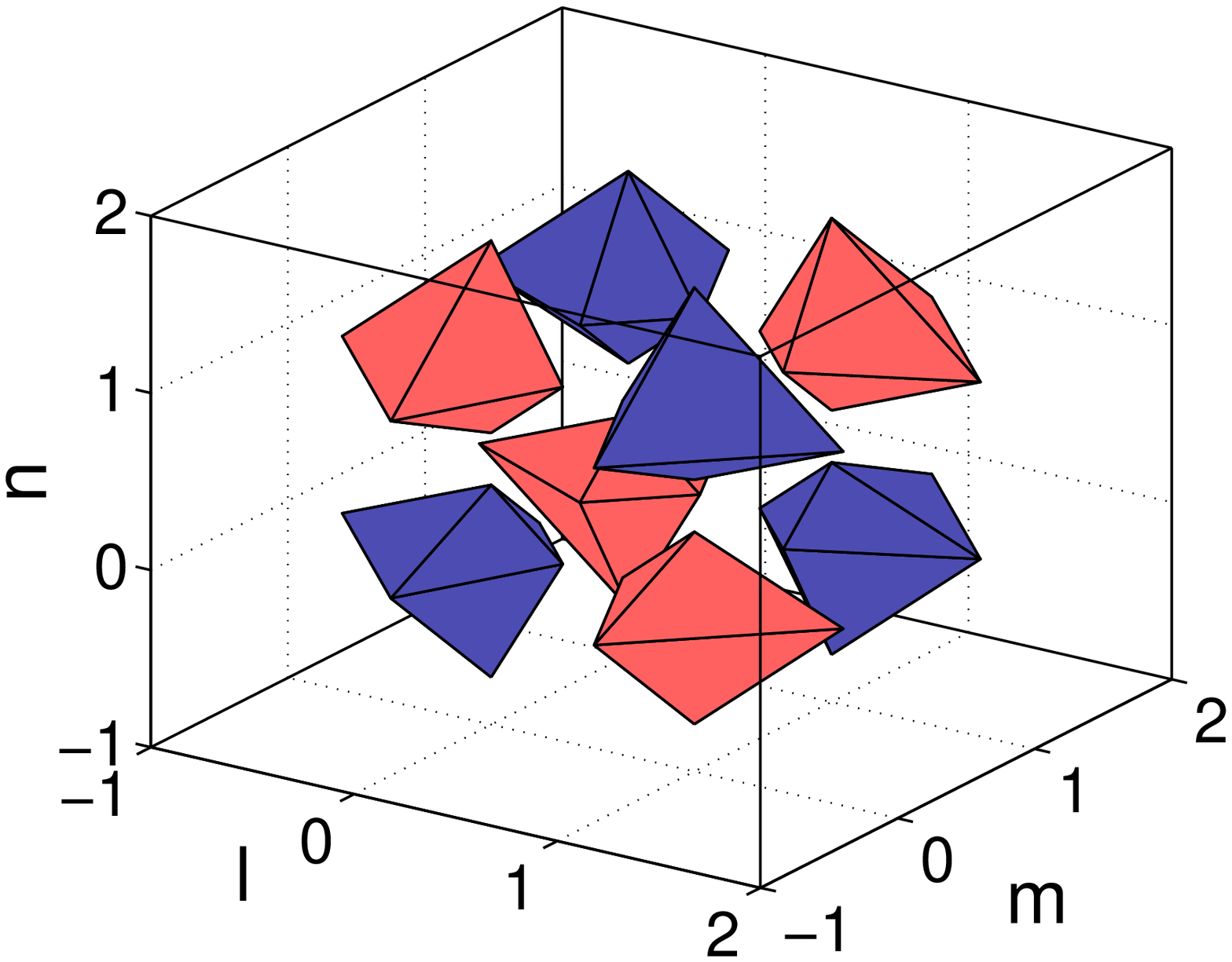} \\[-0.0ex]
\end{tabular}
\begin{tabular}{ll}
(g) & (h) \\[-2.5ex]
\hskip-0.2cm\includegraphics[width=2.95cm]{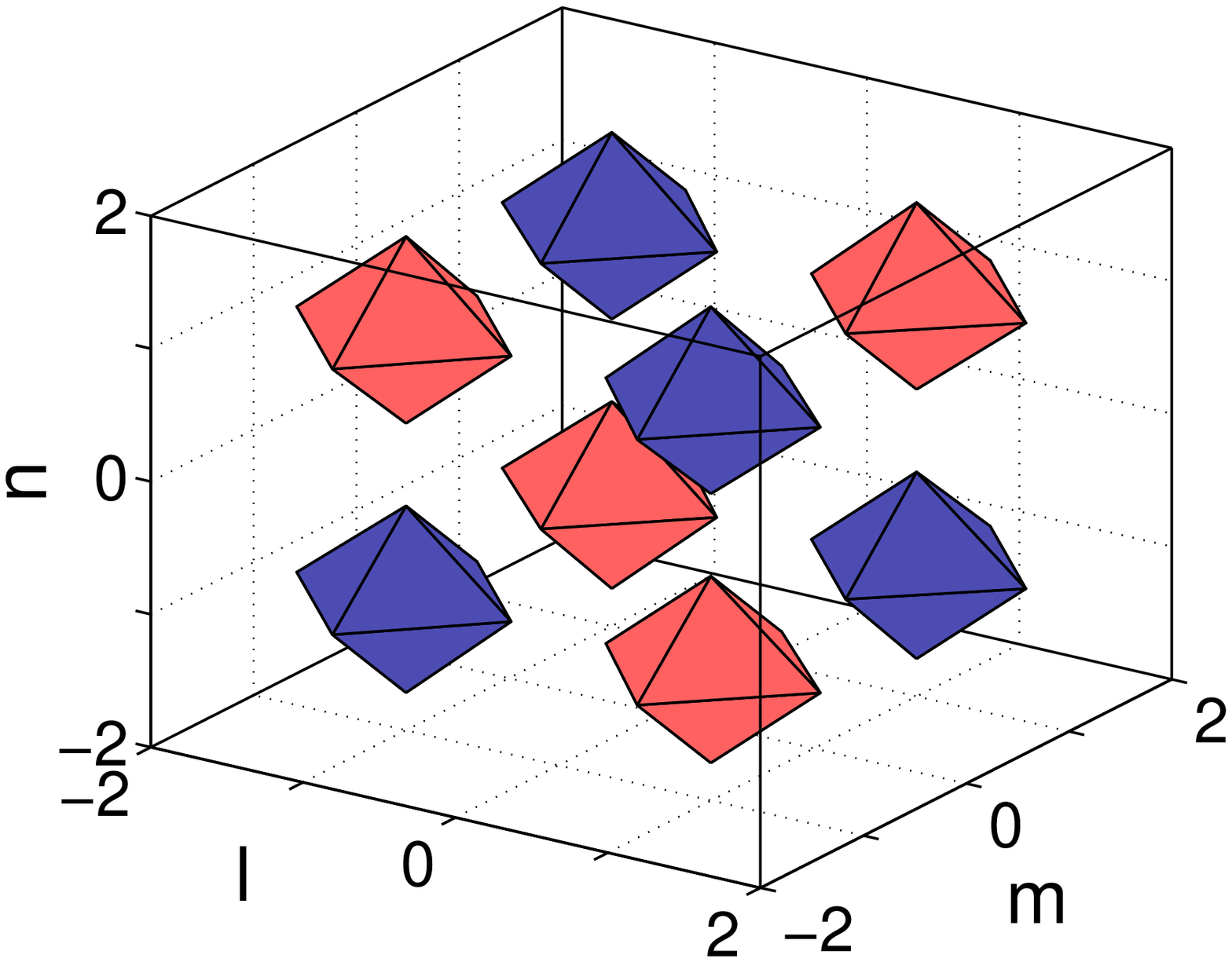} &
\hskip0.0cm\includegraphics[width=5.5cm]{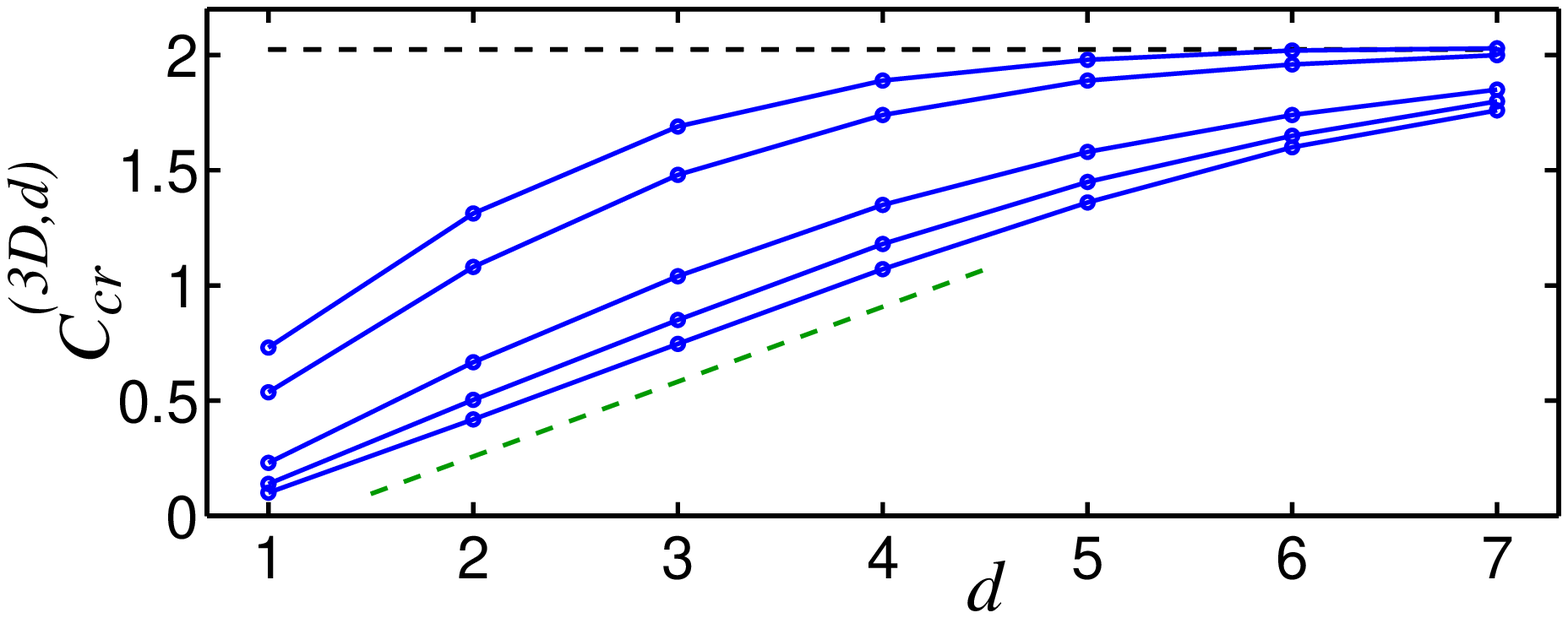} \\[-4.0ex]
\end{tabular}\end{center}
\caption{(Color online) Stable multipoles. The top row depicts
stable ``tight" dipoles (with $d=1$): (a) straight, (b) oblique,
and (c) diagonal ones. (d) and (e): Quadrupoles in the $n=0$
plane, with the internal separation $d=1$ and $d=2$, respectively.
(f) and (g): Octupoles with $d=1$ and $d=2$. Panel (h) displays
the stability threshold
$C_{\mathrm{cr}}^{\mathrm{(3D,}d\mathrm{)}}$ as a function of the
internal distance $d$ for (from top to bottom) diagonal, oblique,
and straight dipoles, octupoles and quadrupoles. The horizontal
dashed line corresponds to the stability threshold for the
fundamental discrete soliton,
$C_{\mathrm{fund}}^{\mathrm{(3D)}}\approx 2$. Note that, for the
quadrupole (the bottom graph),
$C_{\mathrm{quad}}^{\mathrm{(3D,}d\mathrm{)}}$ behaves linearly
for small $d$ (see the dashed line with slope $0.325$ for
guidance). In panels (a)-(g), level contour corresponding to
$\mathrm{Re}(u_{l,m,n})=\pm 0.5$ are shown in blue and red (dark
gray and gray, in the black-and-white version),
respectively. All these states are \emph{stable} (for the case
shown, with $\Lambda =2$ and $C=0.1$).} \label{figs1}
\end{figure}

\textit{Results.} We start by constructing purely real solutions
of the dipole type (in lower-dimensional models, solitons of this type
were considered in Refs.\ \cite{pgk,yang04}). Figure \ref{figs1}
displays generic examples of ``tight" dipoles with adjacent
excited sites (i.e., the separation between them is $d=1$, in the
corresponding units) and three possible orientations relative to
the lattice: \textit{straight} (along a lattice's bond), (a),
\textit{oblique} (along a planar diagonal), (b), and
\textit{diagonal} (along a 3D diagonal), (c). In this figure and
below, unless stated otherwise, we show a typical case admitting
stable solutions, with $C=0.1$ and $\Lambda =2$. The borders of
the stability windows, 
$0\leq C \leq C_{\mathrm{dip}}^{(\mathrm{3D},d)}$, for the three types
(straight, oblique and diagonal) of the dipoles are given by
$C_{\mathrm{dip-str}}^{\mathrm{(3D,1)}}=0.23013\pm \delta C$,
$C_{\mathrm{dip-obl}}^{\mathrm{(3D,1)}}=0.53666\pm \delta C$, and
$C_{\mathrm{dip-dia}}^{\mathrm{(3D,1)}}=0.73084\pm \delta C$,
where the error margin is $\delta C=0.00001$ for all the results
presented in this work (unless indicated otherwise). We observe
that the diagonal dipole in Fig.\ \ref{figs1}(c) remains stable in
a larger interval than its oblique counterparts in Fig.\
\ref{figs1}(b) that, in turn, is more stable that the diagonal one
in Fig.\ \ref{figs1}(c). It is also possible to construct dipole
solutions corresponding to $d=2$, with excited sites separated by
a single nearly empty site. Such dipole solutions (not depicted
here) have larger stability windows than their ``tight" ($d=1$)
counterparts:
$C_{\mathrm{dip-str}}^{\mathrm{(3D,2)}}=0.66722$,
$C_{\mathrm{dip-obl}}^{\mathrm{(3D,2)}}=1.08024$, and
$C_{\mathrm{dip-dia}}^{\mathrm{(3D,2)}}=1.31356$. As seen in Fig.\
\ref{figs1}(h), $C_{\mathrm{dip}}^{\mathrm{(3D},d)}$ further
increases with the separation distance $d$, approaching the
stability threshold of the fundamental (single-site-based)
discrete soliton, $C_{\mathrm{dip}}^{\mathrm{(3D,\infty )}}\equiv
C_{\mathrm{fund}}^{\mathrm{(3D)}}=2.009\pm 0.001$ \cite{DNLS3d},
see the horizontal dashed line in the figure. It can also be found
that the same relations,
$0<C_{\mathrm{dip-str}}^{\mathrm{(3D,d)}}<C_{\mathrm{dip-obl}}^{\mathrm{(3D,d)}}
<C_{\mathrm{dip-dia}}^{\mathrm{(3D,d)}}$,
as found for $d=1$, are valid for all $d$, see Fig.\
\ref{figs1}(h). Another relevant comparison is with dipoles in the
2D DNLS\ model, which were studied in Ref.\ \cite{Bishop}. The
comparison shows that the dipole soliton of the oblique and
straight types, which have their counterparts in the 2D case, are
less stable, although not drastically, than those counterparts:
$C_{\mathrm{dip-str}}^{\mathrm{(3D,1)}}=0.23013<
C_{\mathrm{dip-str}}^{\mathrm{(2D,1)}}=0.245\pm
0.005$ for $\Lambda =2$. The weaker stability of the 3D structures
is explained by the analogy with the multidimensional continuum
NLS equation, where all the solitons are destabilized by collapse,
which is, respectively, \textit{weak} and \textit{strong} in the
2D and 3D cases \cite{sulem}.

Quadrupole and octupole solitons are also shown in Fig.\
\ref{figs1}. The quadrupole is based on four contiguous sites (so
that we prescribe $d=1$ to this structure too) which form a square
in the plane, Fig.\ \ref{figs1}(d). It was found to be stable for
$C<C_{\mathrm{quad}}^{\mathrm{(3D,1)}}=0.13836$, while its 2D
analog has $C_{\mathrm{quad}}^{\mathrm{(2D,1)}}=0.1485\pm 0.0005$
\cite{higher-order2Dvortices}. In this case, as well as it was
with the dipoles, the 2D configurations tend to be slightly more
stable than their 3D siblings. The octupole is shown in panel
\ref{figs1}(f); it is based on a set of eight continuous sites
(therefore it is also assigned $d=1$) forming a cubic cell in the
3D lattice. It is stable in the interval
$C<C_{\mathrm{oct}}^{\mathrm{(3D,1)}}=0.10030$, which is smaller
than the above ones for the quadrupoles and dipoles. Similar to
the dipoles, multipoles can also ``swell" by inserting
unpopulated sites between the excited ones. The resulting
stability intervals for $d=2$ are larger than their $d=1$ counterparts:
$C_{\mathrm{quad}}^{\mathrm{(3D,2)}}=0.503232$ and
$C_{\mathrm{oct}}^{\mathrm{(3D,2)}}=0.418411$, see Figs.\
\ref{figs1}(e) and (g), respectively. It is relevant to mention
that all the newly found structures have their stability limits
lower than those for the fundamental (single-site-based) discrete
soliton, $C_{\mathrm{fund}}^{\mathrm{(3D)}}=2.009\pm 0.001$
\cite{DNLS3d}, see Fig.\ \ref{figs1}(h).

Stable dipoles have one pair of imaginary eigenvalues in their
perturbation-mode spectrum, that, with the increase of $C$, collides with the
continuous spectrum, which leads to the destabilization. 
On the other hand,
stable quadrupoles have three such pairs (two of them form a doublet for
small $C$), and the octupoles have seven (six of which form two triplets for
small $C$). More generally, the number of potentially unstable eigenvalue
pairs is $N-1$, where $N$ is the number of sites on which the structure is
based \cite{pelin}.

\begin{figure}[tbp]
\begin{center}
\begin{tabular}{lll}
(a) & (b) & (c) \\[-2.5ex]
\hskip-0.2cm\includegraphics[width=2.95cm]{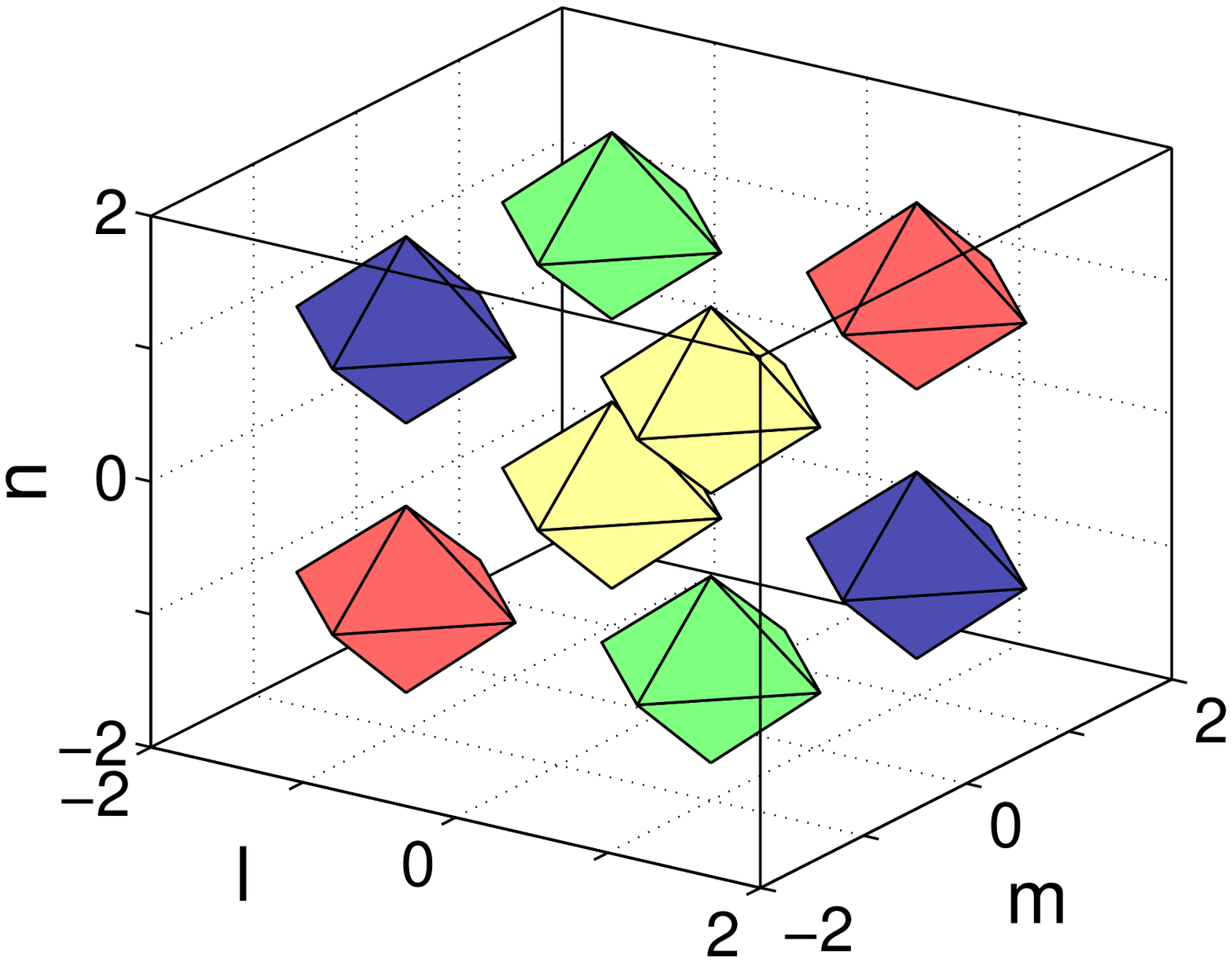} &
\hskip-0.2cm\includegraphics[width=2.95cm]{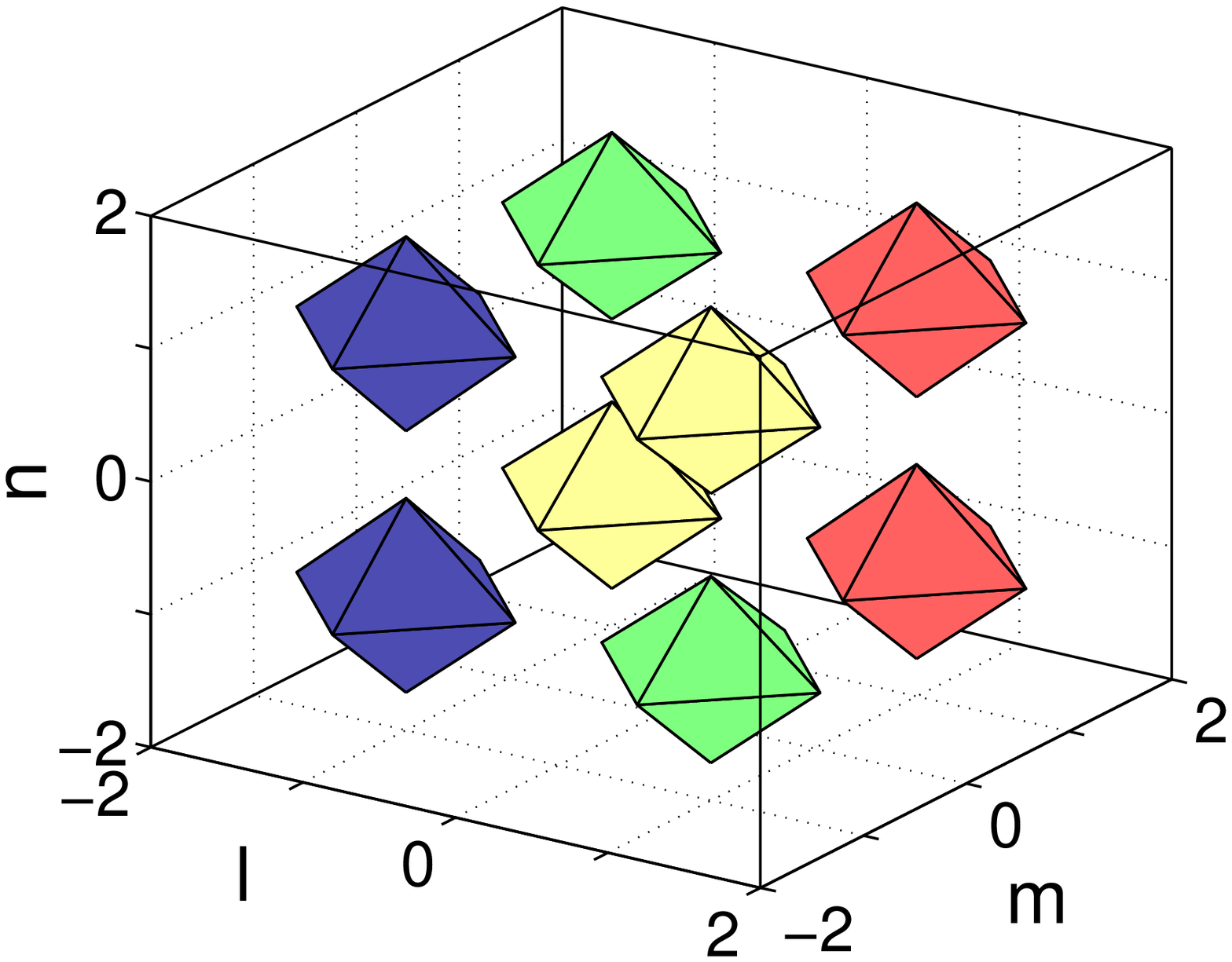} &
\hskip-0.2cm\includegraphics[width=2.95cm]{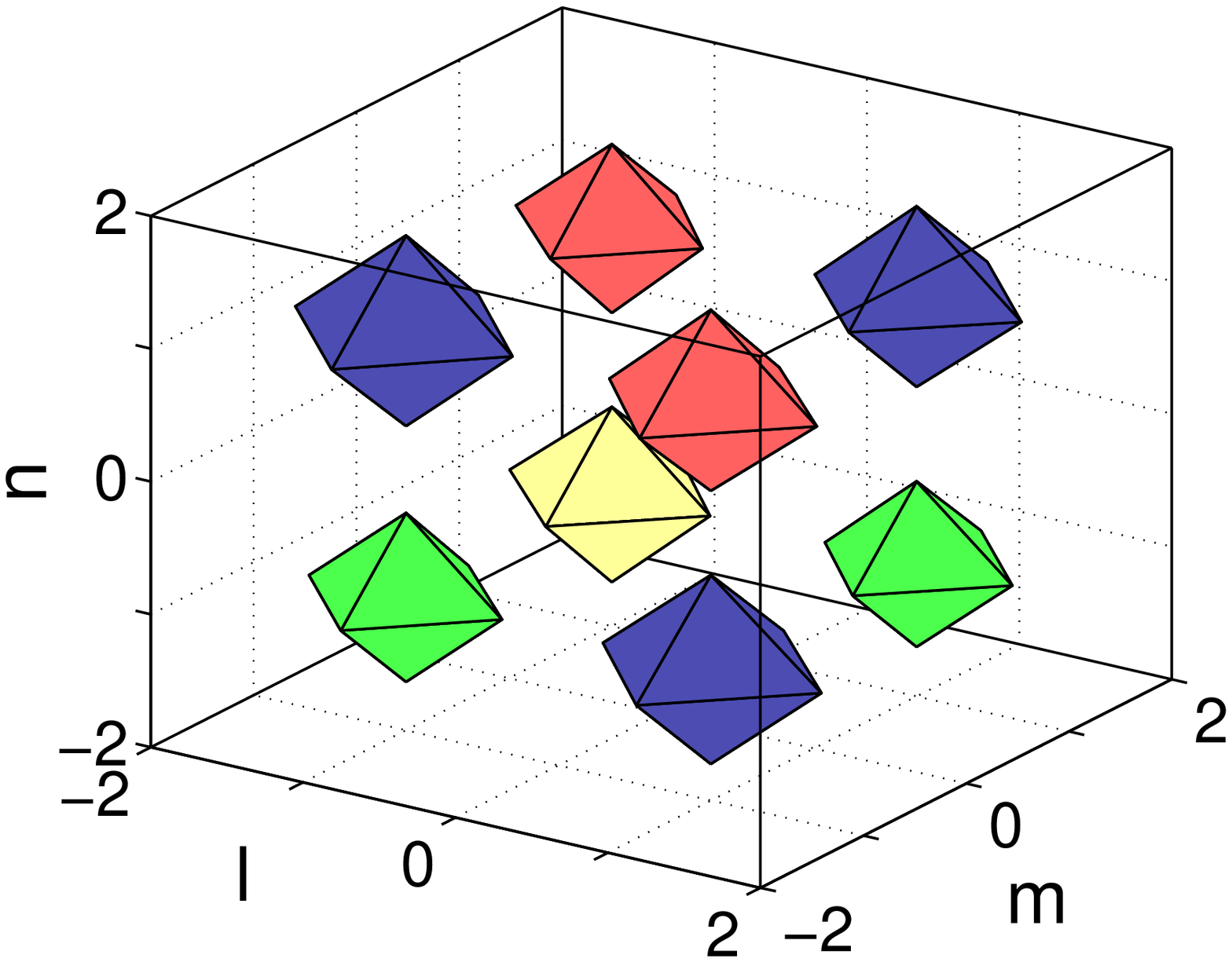} \\[-4.0ex]
\end{tabular}\end{center}
\caption{(Color online) Vortex cubes for $\Lambda =2$ and $C=0.1$.
Panel (a) shows a \emph{stable} vortex cube, built of two
quasi-planar vortices, set in the planes $n=\pm 1$, with equal
vorticities, $S_{1}=S_{2}=1$, and a phase shift of $\protect\pi $.
Panel (b) shows an unstable cube formed by vortices with opposite
charges, $S_{1}=-S_{2}$ and panel (c) shows a snapshot (at
$t=200$) of its evolution, clearly demonstrating that the phase
coherence between sites forming the pattern is lost. The real
level contours are as in Fig.\ \protect\ref{figs1}, and the
imaginary ones, $\mathrm{Im}(u_{l,m,n})=\pm 0.5$, are shown by
green and yellow (light and very light gray, in the
black-and-white version) hues, respectively.} \label{figs2}
\end{figure}


The next novel type of a 3D discrete soliton, with no
lower-dimensional counterpart, is a ``vortex cube", which is built
as a stack of two quasi-planar vortices with equal topological
charges $S_{1}=S_{2}=1$ and a phase shift $\Delta \phi =\pi $,
separated by an empty layer, so that it has $d=2$. Figure
\ref{figs2}(a) shows real and imaginary parts of the vortex-cube
lattice field. Such a state is stable for 
$0\leq C \leq C_{\mathrm{cub,}\pi }^{\mathrm{(3D,2)}}=0.56324$. On the contrary
to this, a vortex cube built as a stack of two in-phase vortices,
with $\Delta \phi =0$, is \emph{always unstable}, through three
real eigenvalue pairs. Further, Fig.\ \ref{figs2}(b) shows a
similar stack, but composed of two vortices with opposite charges,
$S_{1}=-S_{2}=1$. This configuration is \emph{always unstable}
as well, due to a real eigenvalue pair. In this case, the instability
manifests itself as a symmetry breaking between the two planes
and, hence, the phase coherence of the entire pattern is
eventually lost, see Fig.\ \ref{figs2}(c).

\begin{figure}[tbp]
\begin{center}
\begin{tabular}{lll}
(a) & (b) & (c) \\[-2.5ex]
\hskip-0.2cm\includegraphics[width=2.95cm]{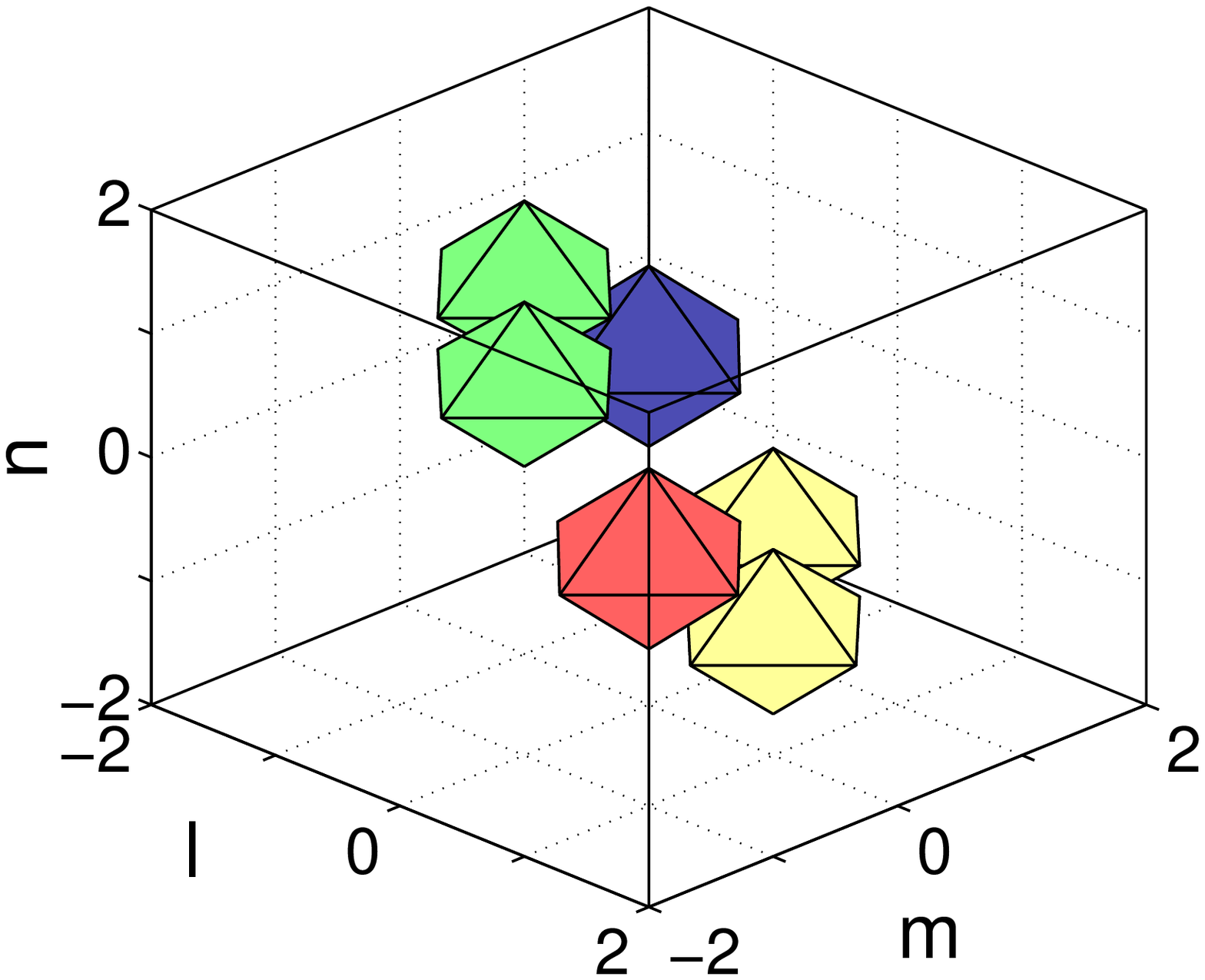} &
\hskip-0.2cm\includegraphics[width=2.95cm]{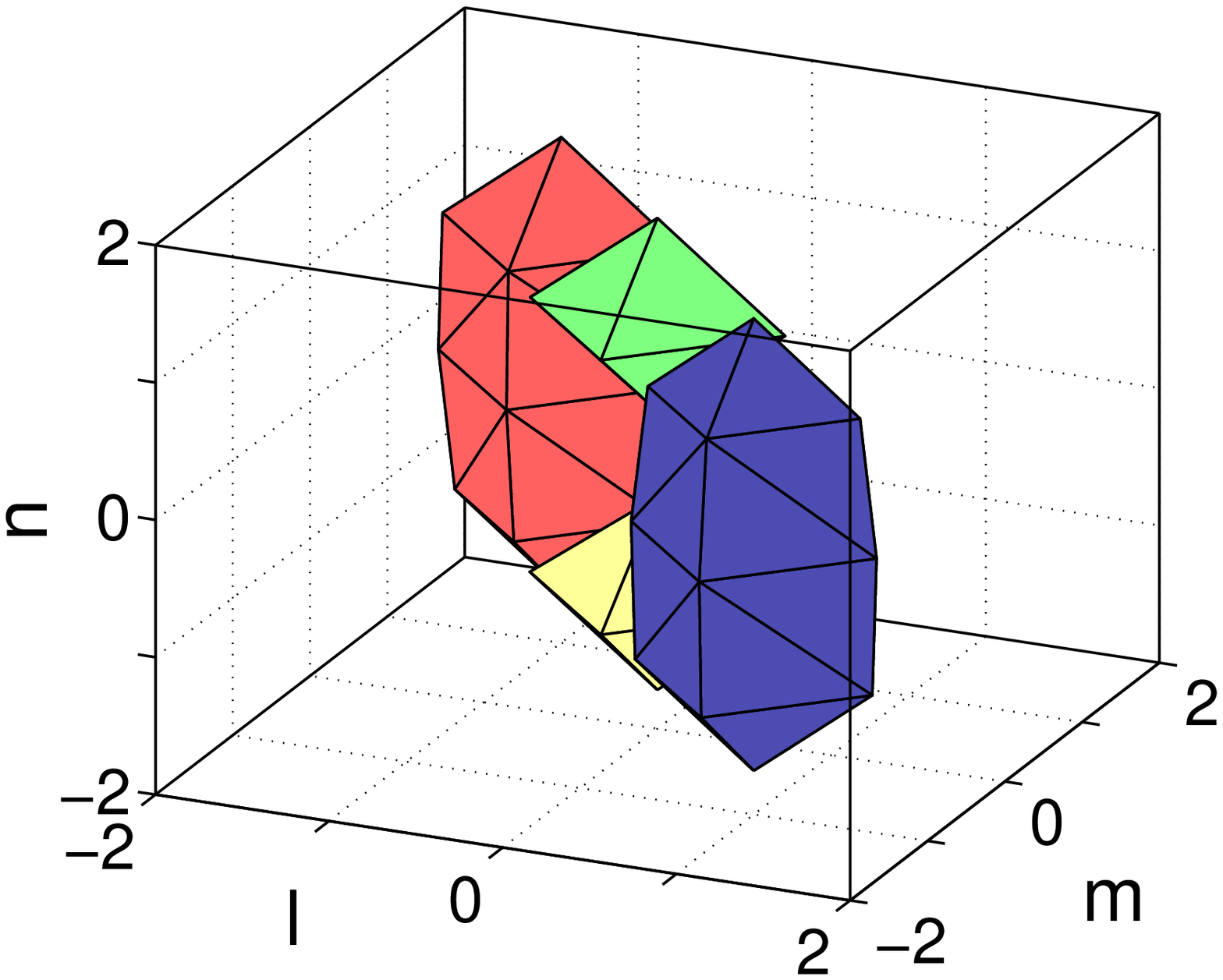} &
\hskip-0.2cm\includegraphics[width=2.95cm]{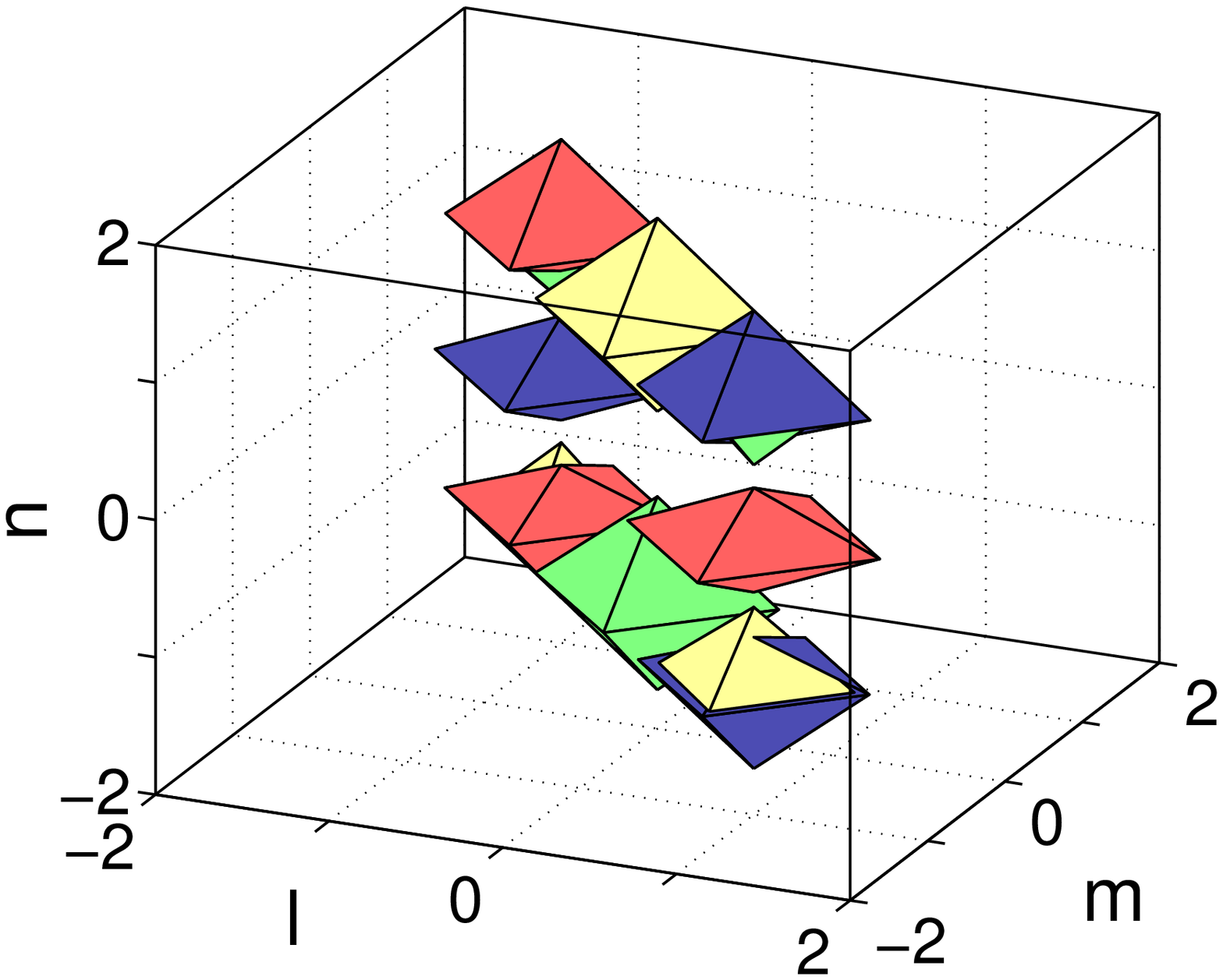} \\[-4.0ex]
\end{tabular}\end{center}
\caption{(Color online) Diagonal and oblique vortices. Panel (a)
shows an unstable ``diagonal" vortex, with the axis along the
direction $(1,1,1)$, for $\Lambda =2$ and $C=0.1$. Panel (b) shows
an unstable oblique vortex with the axis oriented along the
direction $(1,1,0)$, and panel (c) shows a \emph{stable} oblique
vortex of a modified form (see text) for $\Lambda =2$ and $C=0.01$.}
\label{figs3}
\end{figure}

Another 3D object, with no lower-dimensional analog either, is a
vortex with the axis directed along the diagonal of the cubic
lattice, i.e., the vector $(1,1,1)$. Figure \ref{figs3}(a) shows
such a ``diagonal vortex" constructed by a continuation procedure
starting, in the AC limit, with the following distribution of
local phases: $\phi _{l,m,n}=2\pi Sk/6$ ($k=0,1,2,...,5$), for the
sites that lie in a plane orthogonal to the axial (diagonal)
direction: $(1,-1,0)$, $(0,-1,1)$, $(-1,0,1)$, $(-1,1,0)$,
$(0,1,-1)$, $(1,0,-1)$. Obviously, this phase pattern bears the
vorticity $S$ ($S=1$ in Fig.\ \ref{figs3}). However, the diagonal
vortex turns out to be \emph{always unstable}, due to three real
eigenvalue pairs,
and it eventually settles to a single-site-based soliton. 

\begin{figure}[tbp]
\begin{center}
\begin{tabular}{ll}
(a) & \hskip-0.1cm (b) \\[-2.5ex]
\hskip-0.1cm\includegraphics[width=2.95cm]{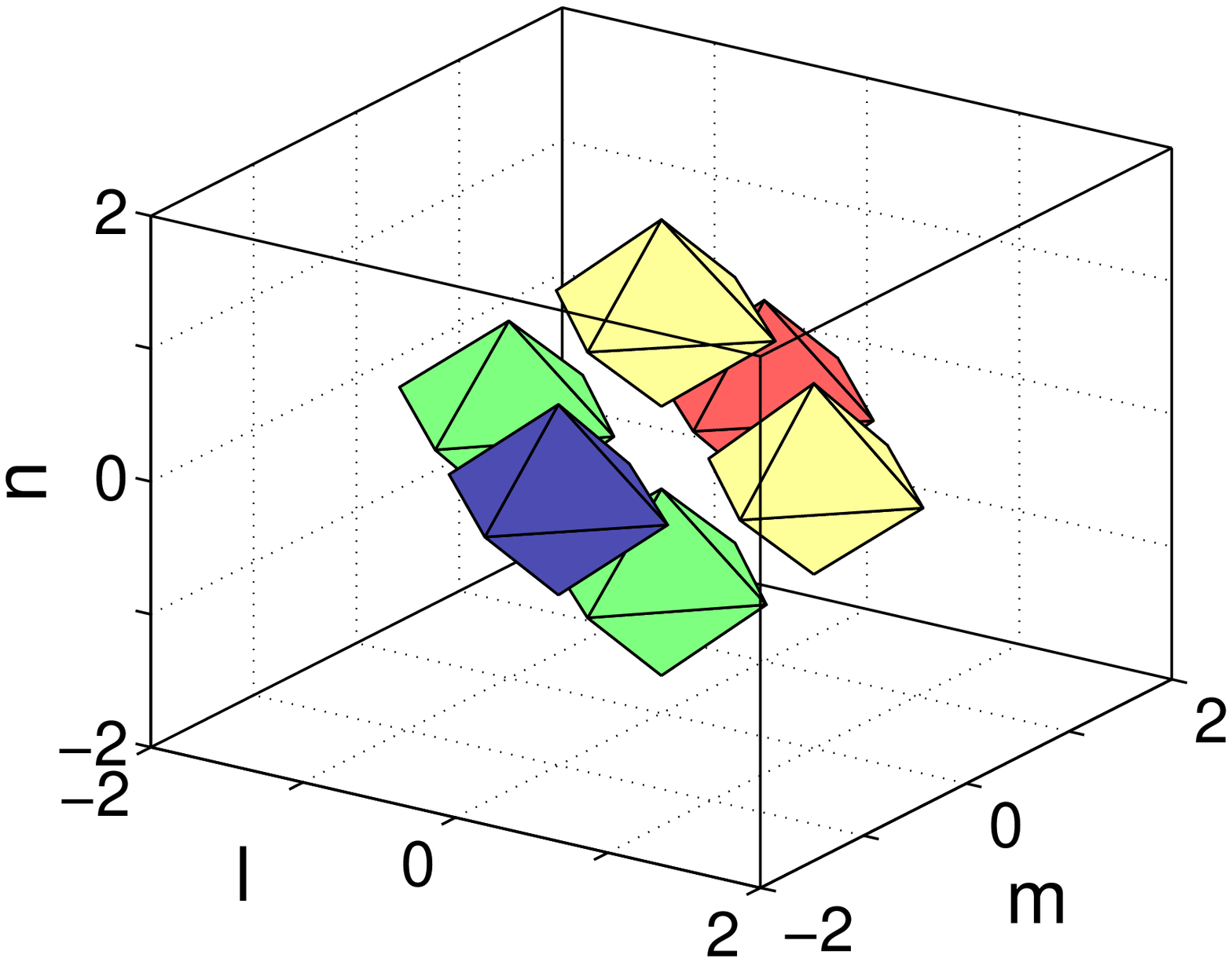} &
\includegraphics[width=5.6cm]{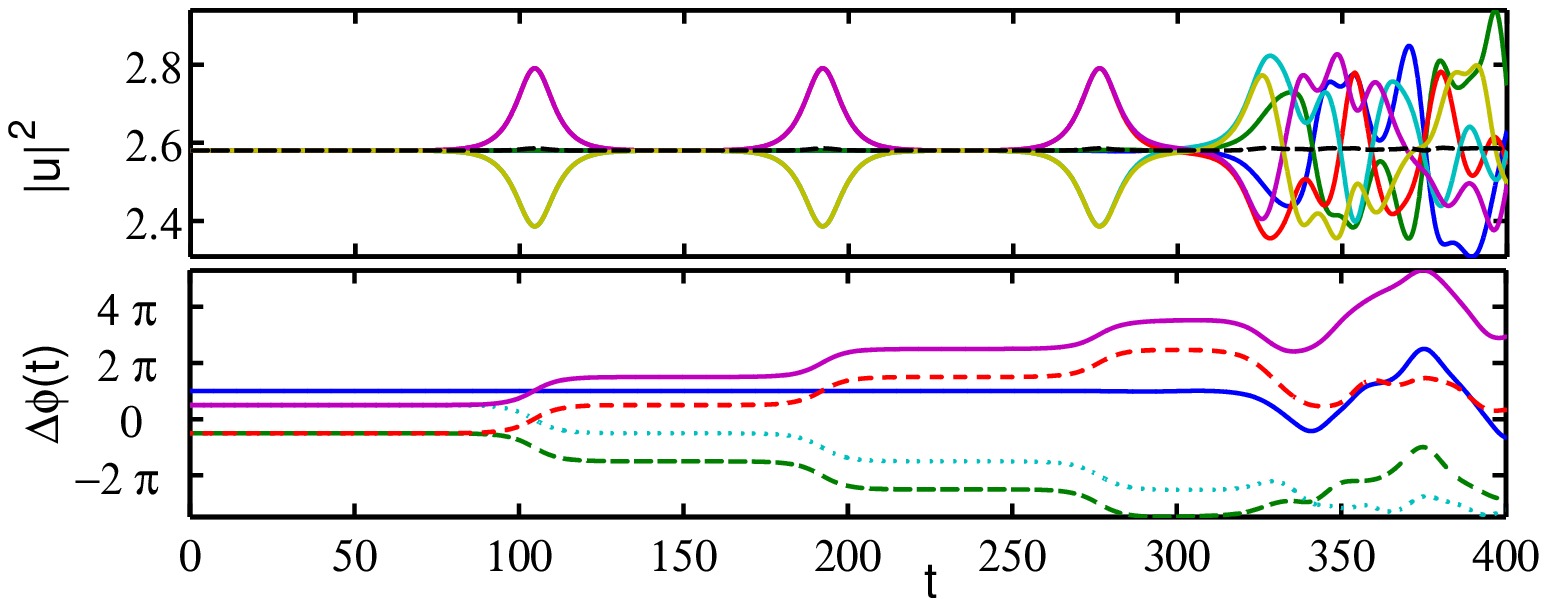} \\[-4.0ex]
\end{tabular}\end{center}
\caption{(Color online) The vortex diamond (discrete skyrmion) for
$\Lambda =2$ and $C=0.1$. Panel (a) displays an unstable
``diamond", and panel (b) depicts its evolution, in terms of the
field's magnitude (top) and phase (bottom) at the main six lattice
sites. Initially ($t<300$), pairs of sites (opposite vertices
of the diamond) cyclically change the phase, and subsequently
($t>350$) the phase coherence is lost and the field magnitude
oscillates erratically about its initial level.} \label{figs4}
\end{figure}

One more species of discrete vortices that may exist solely in the
3D lattice is an ``oblique" one, shown in Figs.\
\ref{figs3}(b)-(c), with the axis' directed along $(1,1,0)$. In
the AC limit, the solution is carried by the array of sites
$(1,-1,0)$, $(1,-1,1)$, $(0,0,1)$, $(-1,1,1)$, $(-1,1,0)$,
$(-1,1,-1)$, $(0,0,-1)$, and $(1,-1,-1)$, with the
phases at them 
$S\cdot (0,\alpha ,\pi /2,\alpha +\pi /2,\pi ,\alpha +\pi ,3\pi
/2,\alpha +3\pi /2)
$, where $\alpha \equiv \tan ^{-1}(1/\sqrt{2})$. 
Figure \ref{figs3}(b) depicts such an oblique vortex, which, in
this form, is found to be \emph{always unstable}, similar to the
diagonal vortex. Nonetheless, the oblique vortex can be stabilized
in a modified form, by introducing a sign shift at the
intermediate edge sites [i.e., $(0,0,1)$, $(-1,1,0)$, $(0,0,-1)$
and $(1,-1,0)$], see Fig.\ \ref{figs3}(c). This sign change avoids
contiguous sites with the same phase and does not alter the
vortex' topological charge, which remains $1$. The modified
oblique
vortex is stable in a small interval, $C<C_{\mathrm{vor-obl}}^{\mathrm{(3D)}}=0.0104$.

Finally, motivated by the concept of skyrmions \cite{skyrme}, which are
distinguished by two topological charges associated with closed
contours in two perpendicular planes, we have constructed one more
type of vortex structures in the 3D lattice, viz., a ``diamond"
shown in Fig.\ \ref{figs4}(a). It is built as a vortex cross,
i.e., a nonlinear superposition of two straight site-centered
$S=1$ vortices, with their axes directed along two orthogonal
directions. The stability analysis shows that, although the
diamond has three imaginary (stable) eigenvalue pairs,
it is \emph{always unstable} due to a real eigenvalue pair. In direct
simulations, its instability manifests itself in a rather intriguing manner,
see Fig.\ \ref{figs4}(b). At $t\approx 100$, two pairs of opposite vertices
of the diamond 
change phases (one by $+\pi /2$ and the other by $-\pi /2$) and
suffer a momentary magnitude change. After this, the diamond
remains stable until $t\approx 200$, when the same pairs of sites
suffer another phase shift (in the same direction). This process
repeats itself almost periodically until (at $t\approx 340$) the
diamond, after 4 shifts, returns back to its original phase
distribution. Subsequently, the phase shifts accelerate and phase
coherence is finally lost. Simultaneously, amplitude variations of
$\pm 10\%$ get accumulated, see the top panel of Fig.\
\ref{figs4}(b). Eventually, the solution degenerates into a plain
single-site-based soliton.

\textit{Conclusions and discussion.} We have introduced
several novel species of topologically structured discrete
solitons in 3D dynamical lattices, using the paradigm 
of the discrete nonlinear Schr{\"{o}}dinger equation. The
solutions have been constructed starting from properly chosen
anti-continuum approximations, and their linear stability was
studied through the computation of the relevant eigenvalues. Previously,
only the fundamental single-site-based solitons and straight
vortices, with the axis directed along a lattice bond, were known.
We have found three species of dipoles, which differ by the
orientation relative to the lattice, quadrupoles and octupoles,
vortex cubes (stacked dual-vortex patterns),
diagonal and oblique vortices, and ``diamonds" (vortex crosses or
discrete skyrmions). Except for the straight and oblique dipoles
and quadrupoles, the patterns obtained are endemic to the 3D lattice
setting, having no counterparts in lower dimensions.

Apart from 
the diagonal vortices and ``diamonds", all the patterns
constructed above have stability regions below a critical value
of the coupling parameter. It is possible to explain the
stability/instability of all the structures that may be realized 
as bound states of two simpler objects, viz., dipoles, quadrupoles
(bound states of two dipoles with opposite orientations),
octupoles (bound states of two quadrupoles), and vortex cubes.
Indeed, a known general principle is that a bound state pinned by
the lattice may be stable only if the coupled objects \emph{repel}
each other \cite{Todd,Bishop} (i.e., have a phase difference of
$\pi$ between their building blocks). This explains the existence of
stability regions for multipoles of all types.
Similarly, considering the interaction between constituent
quasi-planar vortices, one may understand the stability and
instability of vortex cubes of the types shown in Figs.\
\ref{figs2}(a) and \ref{figs2}(b), respectively. Following this
principle, it is also possible to predict the stability of more
exotic 3D patterns such as bound states of two oblique or diagonal
dipoles, or octupoles constructed of two such states. In those
cases when the 3D structures have 2D counterparts, viz., straight
and oblique dipoles and quadrupoles, their stability regions are
narrower than in the 2D case, which is explained by a stronger
trend to collapse in three dimensions. 

Future challenges involve semi-analytical investigation of
such solutions via Lyapunov-Schmidt theory, and identification of 
their stability by means of methods similar to those developed for
the 1D and 2D cases \cite{pelin}. On the experimental side, it may
be interesting to create such structures in BEC loaded into a strong
optical lattice.

\end{document}